\documentclass[preprint]{aastex}

\begin{document}

\title{Physical Nature and Orbital Behavior of the Eclipsing System UZ Leonis}
\author{Jae Woo Lee$^{1,2}$ and Jang-Ho Park$^{1,3}$}
\affil{$^1$Korea Astronomy and Space Science Institute, Daejeon 34055, Korea}
\affil{$^2$Astronomy and Space Science Major, Korea University of Science and Technology, Daejeon 34113, Korea}
\affil{$^3$Department of Astronomy and Space Science, Chungbuk National University, Cheongju 28644, Korea}
\email{jwlee@kasi.re.kr, pooh107162@kasi.re.kr}

\begin{abstract}
New CCD photometric observations of UZ Leo were obtained between February 2012 and April 2013, and on February 2017. 
Its physical properties were derived from detailed analyses of our light curves and existing radial velocities. 
The results indicate that this system is a totally-eclipsing A-subtype overcontact binary with both 
a high fill-out factor of 76 \% and a third light source contributing 12 \% light in the $B$ bandpass, 10 \% in $V$, 
and 7 \% in $R$. The light residuals between observations and theoretical models are satisfactorily fitted by adopting 
a magnetic cool spot on the more massive primary star. Including our 12 measurements, a total of 172 eclipse times 
were used for ephemeris computations. We found that the orbital period of UZ Leo has varied due to a periodic oscillation 
superposed on an upward parabolic variation. The observed period increase at a rate of $+3.49 \times 10^{-7}$ d yr$^{-1}$ 
can be plausibly explained by some combination of non-conservative mass transfer from the secondary to the primary component 
and angular momentum loss due to magnetic braking. The period and semi-amplitude of the oscillation are about 139 yrs and 
0.0225 d, respectively, which is interpreted as a light-time effect due to a third component with a mass of 
$M_3 \sin i_3$ = 0.30 M$_\odot$. Because the third lights of 7$-$12 \% indicate that the circumbinary object is very 
overluminous for its mass, it would possibly match a white dwarf, rather than an M-type main sequence. 
\end{abstract}

{}

\section{INTRODUCTION}

W UMa-type eclipsing binaries consist of two dwarf stars that contact each other physically and are surrounded by a common 
envelope. They have spectral types ranging from late-A to mid-K and are located near or just above the main sequence 
(Bilir et al. 2005). The light curves show continuous brightness variations over time and nearly equal eclipse depths, 
indicating that the effective temperatures of both components are very similar. W UMa contact binaries are divided into 
two subclasses of A and W on the basis of their light curves (Binnendijk 1970). When the larger and more massive component 
is eclipsed by its less massive companion during the primary minimum, it is called A-subtype. On the contrary, 
when the smaller and less massive component is eclipsed at the primary minimum, it is called W-subtype. The formation and 
evolution of these binary systems require a considerable variety of astrophysical processes, but a satisfactory theory for 
them has not been suggested to date (Li et al. 2008; Eggleton 2012).

It is thought that most W UMa stars were formed from detached binaries through orbital decay due to angular momentum loss (AML) 
and ultimately coalesced into single stars (Bradstreet \& Guinan 1994; Eggleton \& Kiseleva-Eggleton 2002). This process could 
only have happened if they were very close binaries to start with and the orbital angular momentum was tidally coupled to 
the spin angular momentum. For the spin-orbit coupling to work efficiently, the initial orbital periods should be several days 
(Bradstreet \& Guinan 1994; Pribulla \& Rucinski 2006). A circumbinary object may play an important role in the formation 
of an initial tidal-locked detached binary by transferring angular momentum via the combined action of Kozai cycle (Kozai 1962) 
and tidal friction (Fabrycky \& Tremaine 2007). Tokovinin et al. (2006) found that almost all close binaries with periods shorter 
than 3 d exist in multiple systems. The statistical study of Pribulla \& Rucinski (2006) indicated that a large proportion of 
the W UMa systems have circumbinary companions, which are necessary for the origin and evolutionary processes of contact binaries. 
The presence of a third companion can be inferred from the analyses of both light curves and eclipse timings. 

UZ Leo ($\rm BD+14^{o} 2280$, HIP 52249, TYC 845-995-1; $V\rm_T$=$+$9.80, $(B-V)\rm_T$=$+$0.41) is an A-subtype W UMa eclipsing 
binary with an orbital period of about 0.6180 d. It was discovered as a cluster-type variable by Kaho (1937), and it was then 
considered as an RR Lyr-type star by many other investigators (e.g., Ashbrook 1949; K\"uhn 1951). The correct W UMa type of 
the variable was identified by the photoelectric observations of Smith (1954, 1959). The full light curves of UZ Leo were 
obtained by Binnendijk (1972), Heged\"us \& J\"ager (1992), and Zola et al. (2010). Rucinski \& Lu (1999; hereafter RL) presented 
double-line radial-velocity (RV) curves with semi-amplitudes of $K_1$ = 76.8 km s$^{-1}$ and $K_2$ = 262.9 km s$^{-1}$, and 
classified the spectral type of this star to be A9$-$F1V. Most recently, Zola et al. (2010) analyzed their $BVR$ light curves 
and determined the absolute dimensions from the photometric parameters and the spectroscopic orbit of RL. These indicate that 
UZ Leo is an overcontact binary with an orbital inclination of $i$ = 86$^\circ$.6, a temperature difference of $\Delta T$ 
= 150 K, a fill-out factor of $f$ = 97 \%, and third lights of $\ell_3$ = 10$-$14 \% in three bandpasses. On the other hand, 
the orbital period of the eclipsing system has been studied by Heged\"us \& J\"ager (1992) and Zola et al. (2010), who showed 
that the period is increasing. They reported that the eclipse timing variation may be the result of mass transfer from the less 
massive secondary to the more massive primary component. The latter authors also suggested that the secular period increase 
could be a small part of a larger, cyclic curve caused by the presence of a third body orbiting the binary system. 
In the remainder of this paper, we present and discuss the physical nature and orbital behavior of UZ Leo based on detailed 
studies of our new CCD observations as well as historical data.

\section{CCD PHOTOMETRIC OBSERVATIONS}

We performed new CCD photometry of UZ Leo during two observing seasons between February 2012 and April 2013, using 
a PIXIS: 2048B CCD and a $BVR$ filter set attached to the 61-cm reflector at Sobaeksan Optical Astronomy Observatory (SOAO) 
in Korea. The CCD chip has 2048$\times$2048 pixels and a pixel size of 13.5 $\mu$m, so the field of view of a CCD frame is 
17$\arcmin$.6$\times$17\arcmin.6. With the conventional IRAF package, we processed the CCD frames to correct for bias level 
and pixel-to-pixel inhomogeneity of quantum efficiency, and applied simple aperture photometry to obtain 
instrumental magnitudes for tens of stars imaged on the chip at the same time as the variable. TYC 845-996-1 
(GSC 0845-0996; $V\rm_T$=$+$12.68, $(B-V)\rm_T$=$+$0.81) was used as a comparison star, and no brightness variation of it 
was detected against measurements of the other stars monitored during our observing runs.

From the time-series photometry, we obtained 3733, 3722, and 3686 individual points in the $B$, $V$, and $R$ bandpasses, 
respectively, and a sample of them is presented in Table 1. The natural-system light curves of UZ Leo are plotted in Figure 1 
as differential magnitudes {\it versus} orbital phases, which were computed according to the ephemeris for our spot model, 
which is described in section 4. The open circles and plus symbols represent the individual measurements of the 2012 and 
2013 seasons, respectively. In addition to these complete light curves, two primary eclipses were observed on February 2017 
using a FLI 4K CCD and a $R$ filter attached to the same telescope. TYC 845-996-1 also served as a comparison star for 
these data collections.

\section{ECLIPSE TIMING VARIATIONS}

From all SOAO observations, 12 new eclipse times and their errors were determined with the weighted means for the timings 
obtained in each filter by using the method of Kwee \& van Woerden (1956). In addition, we derived two minimum epochs 
from the data of Zola et al. (2006). A total of 172 archival timings (51 visual, 16 photographic plate, 33 photoelectric, 
and 72 CCD) were collected from the literature and from our measurements. Most of the earlier timings were extracted 
from the data base published by Kreiner et al. (2001). All photoelectric and CCD timings are listed in Table 2, 
where the second column gives the error of each minimum determination. For the period study of UZ Leo, we computed 
the 1$\sigma$-values of the scatter bands of the timing residuals to provide mean errors for the observational methods 
as follows: $\pm$0.022 d for visual, $\pm$0.0091 d for photographic, $\pm$0.0027 d for photoelectric, and $\pm$0.0012 d 
for CCD minima. Relative weights were then scaled from the inverse squares of these values.

As mentioned in the Introduction, previous researchers (Heged\"us \& J\"ager 1992; Zola et al. 2010) reported that the orbital 
period change of UZ Leo can be represented by a parabola caused by a mass transfer between the components. As an alternative, 
Zola et al. (2010) suggested that the parabolic variation may be only a part of a period modulation due to the presence of 
a circumbinary companion on a wide orbit. After testing possible ephemeris models, we knew that the eclipse timings would be 
best described by a sinusoidal variation superposed on an upward parabola, rather than by a monotonic way. The sinusoidal term 
was tentatively considered as a light-travel-time (LTT) effect. As a result, the timing residuals were finally fitted to 
the following quadratic {\it plus} LTT ephemeris (e.g., Lee et al. 2015a,b):
\begin{eqnarray}
C = T_0 + PE + AE^2 + \tau_{3}, 
\end{eqnarray}
where $\tau_{3}$ is the LTT due to a third object orbiting the eclipsing binary (Irwin 1952, 1959), and it includes five 
parameters ($a_{\rm b}\sin i_3$, $e_{\rm b}$, $\omega_{\rm b}$, $n_{\rm b}$, and $T_{\rm b}$). 

The Levenberg-Marquart technique (Press et al. 1992) was applied to solve for the eight parameters of equation (1), and 
the results are summarized in Table 3. In this table, $P_{\rm b}$ and $K_{\rm b}$ denote the cycle length and semi-amplitude 
of the LTT orbit, respectively. The absolute dimensions presented in the following section have been used for these and 
subsequent calculations. The eclipse timing diagram of UZ Leo is plotted in the top panel of Figure 2, where the solid curve 
and the dashed parabola represent the full contribution and the quadratic term, respectively. The middle panel refers to 
the LTT orbit, and the bottom panel shows the photoelectric and CCD residuals from the full ephemeris. These appear as 
$O$--$C_{\rm full}$ in the fourth column of Table 2. The quadratic {\it plus} LTT ephemeris resulted in a smaller reduced 
$\chi^2_{\rm red}$=1.075 than either a parabola ($\chi^2_{\rm red}$ = 1.214) or a single LTT ($\chi^2_{\rm red}$ = 1.118). 
As seen in Figure 2, our ephemeris model gives a satisfactory fit to the timing residuals. The mass function of 
the tertiary companion becomes $f_3 (M)$ = 0.0032 $M_\odot$, and its projected mass is $M_{3} \sin i_{3}$ = 0.30 $M_\odot$.

\section{LIGHT-CURVE SYNTHESIS AND ABSOLUTE DIMENSIONS}

Like historical data, our observations presented in Figure 1 display a typical light curve of a contact binary and a flat bottom 
at the secondary minimum. These indicate that the smaller and cooler secondary component is totally occulted by the primary 
component and that UZ Leo belongs to the A-subtype of W UMa-type stars. To obtain a consistent set of binary parameters, 
we analyzed the SOAO light curves with the RV measures of RL by using contact mode 3 of the Wilson-Devinney synthesis code 
(Wilson \& Devinney 1971; van Hamme \& Wilson 2007; hereafter W-D). In this article, the subscripts 1 and 2 refer to 
the primary and secondary components being eclipsed at orbital phases 0.0 (Min I) and 0.5 (Min II), respectively. 

The light and RV curves of UZ Leo were modeled in a manner analogous to that for the contact binaries V407 Peg (Lee et al. 2014) 
and DK Cyg (Lee et al. 2015b). The surface temperature of the primary star was set to be $T_{1}$ = 6,980$\pm$250 K, according 
to the spectral type A9$-$F1V classified by RL and Harmanec's (1988) table. The bolometric ($X$, $Y$) and bandpass ($x$, $y$) 
limb-darkening coefficients for the logarithmic law were interpolated from the values of van Hamme (1993). The initial values 
for most parameters ($i$, $T_{1,2}$, $\Omega_1$, and $L_1$) were adopted from Zola et al. (2010). Adjustable parameters were 
the orbital ephemeris ($T_0$ and $P$), the system velocity ($\gamma$), the semi-major axis ($a$), the mass ratio ($q$), 
the orbital inclination ($i$), the temperature ($T_2$) of the secondary star, the dimensionless surface potentials 
($\Omega_1$ = $\Omega_2$), the monochromatic luminosity ($L_1$), and the third light ($\ell_3$). 

Although the simultaneous analysis of light and RV curves is possible, it is hard to allocate appropriate weights to them 
if there is starspot variability with time and the light curves are not made contemporaneously with the RVs. Thus, 
we satisfactorily analyzed different types of observations through two steps. First, the SOAO light curves alone were 
modeled with a spectroscopic mass ratio. Then, the RV curves of RL were solved using the photometric parameters obtained 
in the first step. This process was iterated until the results for the two data sets were consistent with each other. 
The unspotted solution is listed in the second and third columns of Table 4 and appears as the dashed curves in Figure 3. 
The residuals from the analysis were computed to see the details of unmodeled lights, and they are shown in the left panels 
of Figure 4, where they display quasi-periodic variation patterns. This feature could be attributed to the spot activity on 
the stellar photosphere. 

To explain the residual light discrepancy from the unspotted model, we reanalyzed the light and RV curves by adding a starspot 
on the more massive primary component (Mullan 1975). Although the short- and long-term light variations of many W UMa stars 
may be caused by the changes in the spot parameters with time (e.g., Lee et al. 2010), the differences between the two seasonal 
light curves of UZ Leo are insignificant. Thus, we simultaneously modeled all SOAO data as before. The best result for 
this model is given in columns (4)$-$(5) of Table 4 together with the spot parameters. The synthetic light curves are plotted 
as the solid curves in Figure 3, and the synthetic RV curves are plotted in Figure 5. The residuals from the cool-spot model 
are plotted in the right panels of Figure 4. From these figures, we can see that the single spot model is sufficient to fit 
the quasi-sinusoidal light variations. Our light-curve synthesis indicates that UZ Leo is a totally-eclipsing A-subtype 
overcontact binary with both a high filling factor of about 76 \% and a temperature difference of 208 K between the components, 
and that $\ell_3$ contributes 7$-$12 \% light in all bandpasses. 

Our light-curve synthesis provided a good fit to the light and RV curves, and allowed us to compute the fundamental parameters 
of UZ Leo listed in Table 5. The luminosity ($L$) and bolometric magnitudes ($M_{\rm bol}$) were computed by adopting 
$T_{\rm eff}$$_\odot$ = 5,780 K and $M_{\rm bol}$$_\odot$ = +4.73. The bolometric corrections (BCs) were taken from 
the correlation between $\log T_{\rm eff}$ and BC recalculated by Torres (2010). With an apparent visual magnitude of 
$V$ = +9.75 at maximum light (Deb \& Singh 2011), our computed light ratio, and the interstellar absorption of 
$A_{\rm V}$=0.10 (Schlafly \& Finkbeiner 2011), we determined the distance of the system to be 388 $\pm$ 29 pc. This is 
too large compared with the value 150$\pm$30 pc taken by trigonometric parallax (6.65$\pm$1.34 mas) from the Hipparcos and 
Tycho Catalogues (ESA 1997), while it is consistent with 452$\pm$115 pc computed from Gaia DR1 (2.21$\pm$0.56 mas; 
Gaia Collaboration et al. 2016) within their errors.

\section{DISCUSSION AND CONCLUSIONS}

In this article, we presented and analyzed new CCD observations of UZ Leo, together with historical data collected from 
the literature. The light curves display a total eclipse at secondary minimum and a quasi-sinusoidal light variation, which 
are modeled by using a cool-spot model on the primary component and considering a third light ($\ell_3$) to the system. 
Because UZ Leo should be a fast-rotating binary with a common convective envelope, it is reasonable to regard the starspot 
as a result of magnetic dynamo-related activity. From the combined light and RV solution, we found that the absolute dimensions 
of both components are $M_1$ = 2.01 M$_\odot$, $M_2$ = 0.62 M$_\odot$, $R_1$ = 2.23 R$_\odot$, and $R_2$ = 1.40 R$_\odot$, 
$L_1$ = 10.6 L$_\odot$, and $L_2$ = 3.68 L$_\odot$. A comparison of the UZ Leo parameters with the mass-radius, 
mass-luminosity, and Hertzsprung-Russell (HR) diagrams (Lee et al. 2014) shows that the hotter and more massive primary lies 
in the main-sequence band, and the low-mass secondary is oversized and overluminous for its present mass. This indicates that 
a large amount of energy might be transferred from the primary component to the less massive secondary (Lucy 1968; Li et al. 2008). 

The orbital period of UZ Leo is thought to have changed by a combination of an upward parabola and a periodic variation 
with a cycle length of 139 yr and a semi-amplitude of 0.0225 d. The positive coefficient of the quadratic term ($A$) in 
Table 3 indicates a continuous period increase with a rate of $P/dt = +3.49 \times 10^{-7}$ d yr$^{-1}$. Because UZ Leo is 
a W UMa-type overcontact binary system, the parabolic change can be produced by a mass transfer from the secondary component 
to its more massive primary. Under the assumption of conservative mass transfer, we obtained a modest rate of 
1.69$\times$10$^{-7}$ M$_\odot$ yr$^{-1}$. If the secondary star transfers its present mass to the primary component on 
a thermal time scale $\tau_{\rm th}$ = $(GM_{2}^2)/(R_{2}L_{2})$ = 2.33$\times 10^{6}$ yr, the predicted rate is 
$M_{\rm 2}/ \tau_{\rm th}$ = 2.66$\times 10^{-7}$ M$_{\sun}$ yr$^{-1}$. This value becomes about 160 \% of the observed 
rate deduced from our timing analysis, which indicates that the mass-transfer process in the binary is not conservative. 
Thus, the possible explanation of the parabolic term may be some combination of non-conservative mass transfer and AML 
due to magnetic stellar wind braking. 

The periodic component in the eclipse timing variation could be explained as the LTT effect driven by a circumbinary object 
in this system. If a third companion is on the main sequence and its orbit is coplanar with the eclipsing binary 
($i_3 \simeq$ 87$^\circ$), the mass and radius of this object are computed to be $M_3$ = 0.30 M$_\odot$ and $R_3$=0.31 R$_\odot$, 
respectively, following the empirical mass-radius relation of Southworth (2009). Because these values correspond to 
a low-mass M-type dwarf, the tertiary body would contribute about 1 \% to the total light of the triple system. However, 
our light-curve solution reveals $\ell_3$ to contribute 12 \% light in the $B$ bandpass, 10 \% in $V$, and 7 \% in $R$. 
Assuming that the third lights come from the circumbinary object, its color index can be estimated to be ($B-V$)$_3 \simeq$ 
$+$0.07 (Borkovits et al. 2002), which corresponds to a spectral type of about A2. Hence, if it really exists, the circumbinary 
object has to be very overluminous for its mass, which may be a white dwarf. Alternatively, the period modulation could be 
produced by a magnetic activity cycle of a solar-type component (Applegate 1992; Lanza et al. 1998). However, this mechanism 
displays quasi-sinusoidal timing variations for systems with spectra later than F5 (Hall 1989, Liao \& Qian 2010), 
unlike our program target UZ Leo. Further, it is not easy for the Applegate model to produce a perfectly smooth timing variation. 
Accordingly, the periodic oscillation most likely arises from the LTT effect due to a third companion. The circumbinary object 
is the main source of the third lights detected in our light curves and provides a significant clue to the formation and 
evolution of the eclipsing pair. A large number of future accurate timings are needed to identify the presence of 
the supposed third component and to reveal more detailed properties of the stellar system.

\acknowledgments{ }

The authors would like to thank the staff of SOAO for assistance during our observations. We also thank Professor Chun-Hwey Kim 
for his help using the $O$--$C$ database of eclipsing binaries. We have used the Simbad database maintained at CDS, Strasbourg, 
France. This work was supported by the KASI grant 2017-1-830-03.

\clearpage
\begin{figure}
\includegraphics[]{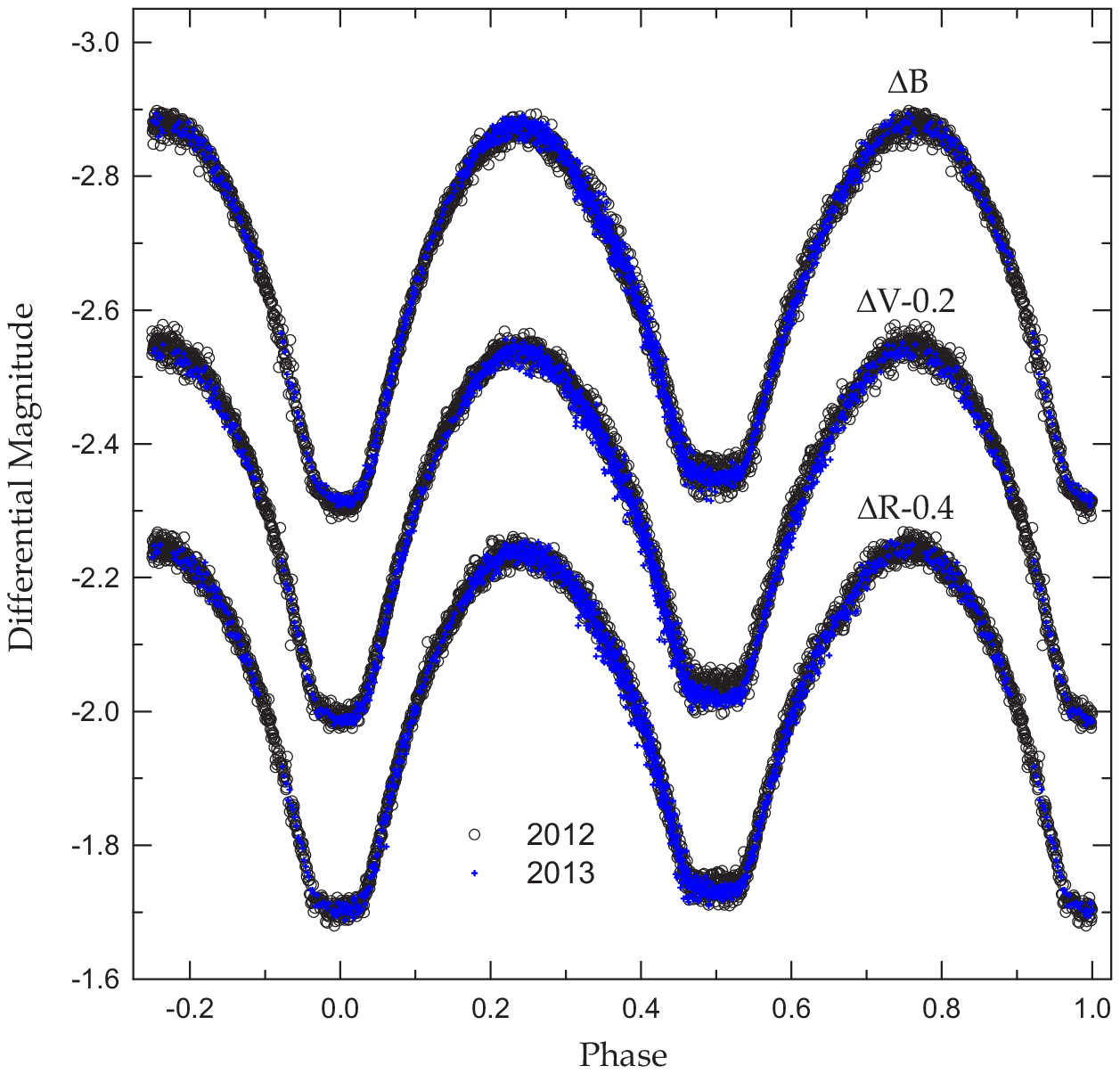}
\caption{$BVR$ light curves of UZ Leo observed at SOAO. The open and plus symbols represent the individual measurements of 
the 2012 and 2013 seasons, respectively. }
\label{Fig1}
\end{figure}

\begin{figure}
\includegraphics[]{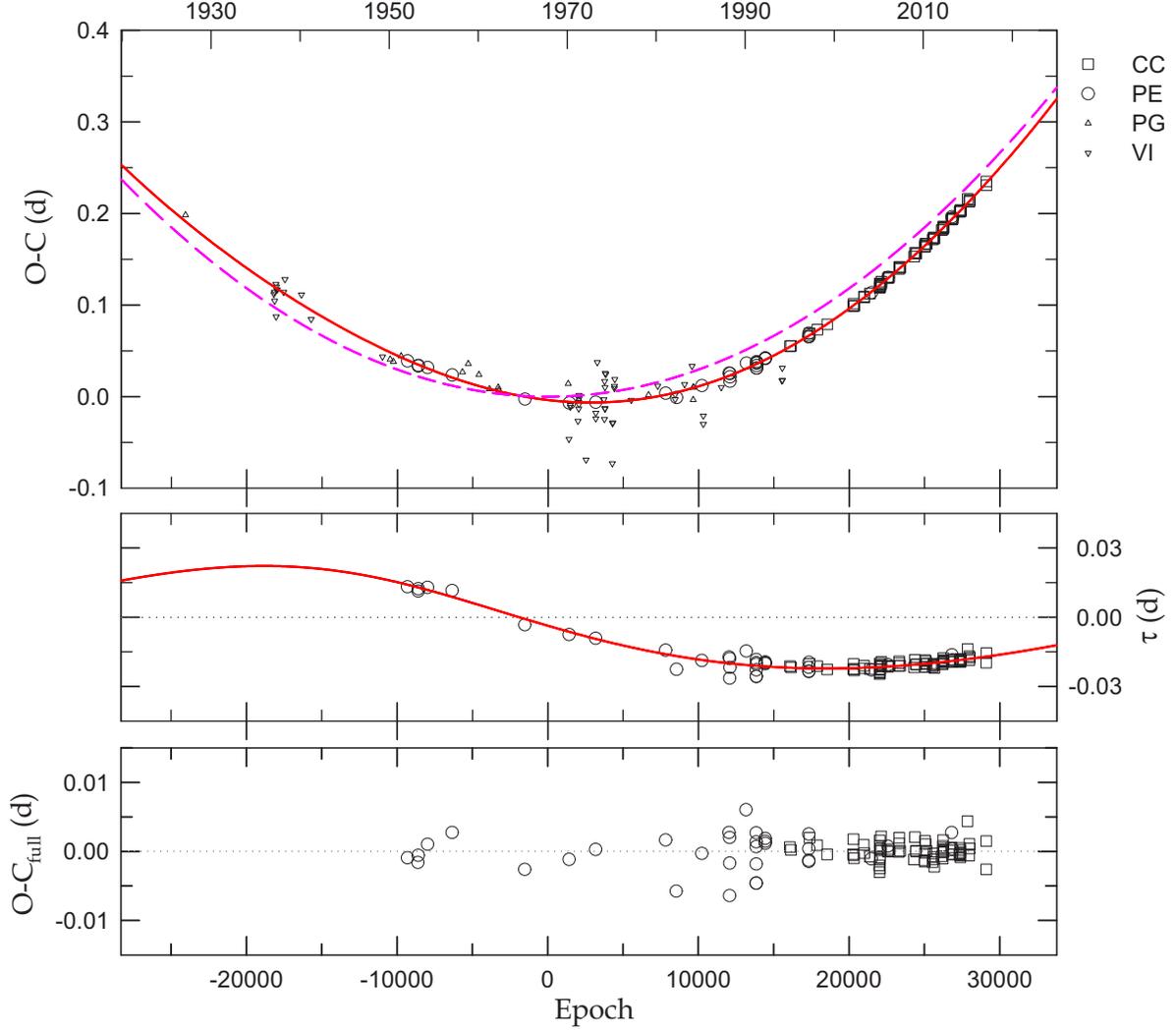}
\caption{Eclipse timing $O$--$C$ diagram of UZ Leo with respect to the linear terms in Table 3. In the top panel, the solid 
and dashed curves represent the full contribution and the quadratic term of equation (1), respectively. The middle panel refers 
to the LTT orbit ($\tau_3$), and the bottom panel shows the photoelectric and CCD residuals from the complete ephemeris. CC, PE, 
PG, and VI denote CCD, photoelectric, photographic, and visual minimum epochs, respectively. }
 \label{Fig2}
\end{figure}

\begin{figure}
\includegraphics[]{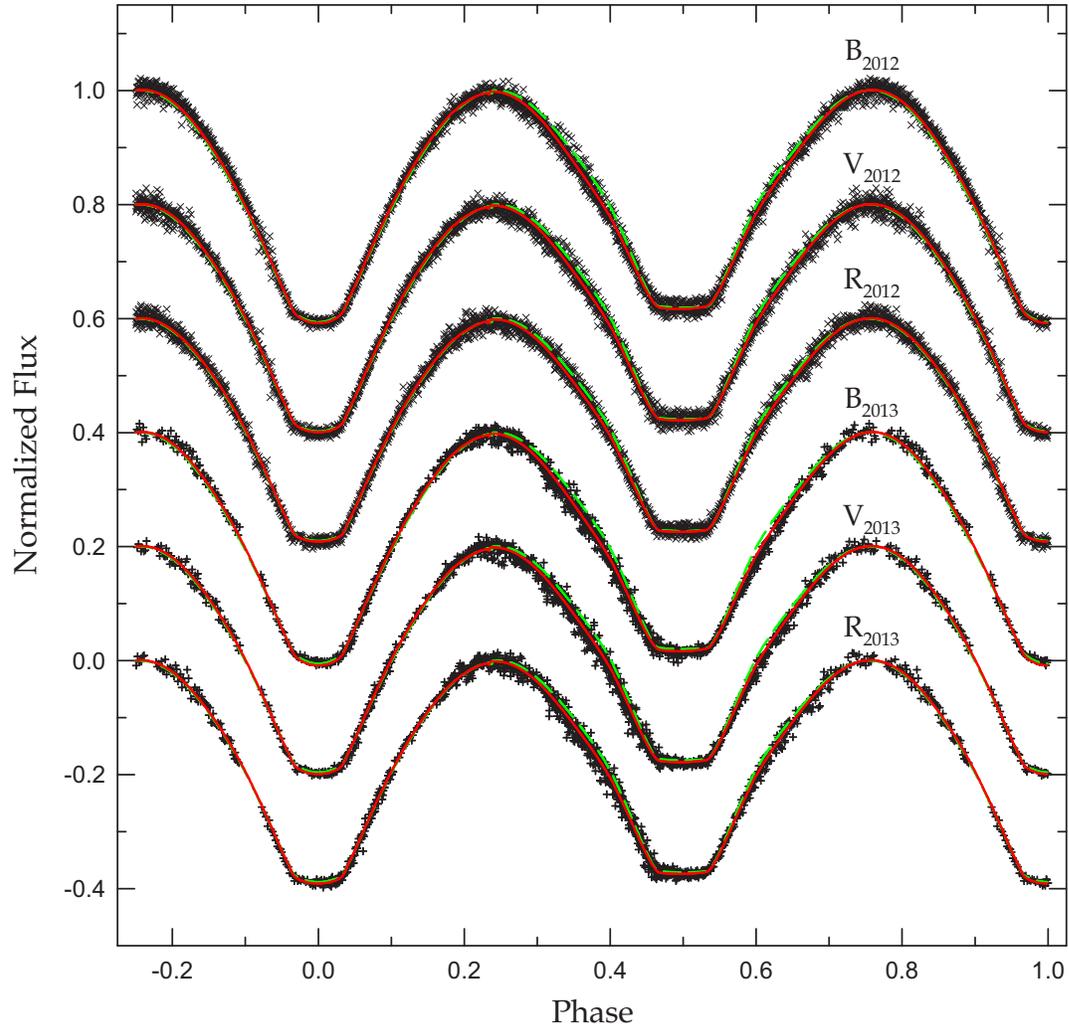}
\caption{Normalized observations with the fitted model light curves. The dashed and solid curves represent the solutions 
obtained without and with a spot, respectively, listed in Table 4. All but $B_{2012}$ are displaced vertically for clarity. }
\label{Fig3}
\end{figure}

\begin{figure}
\includegraphics[]{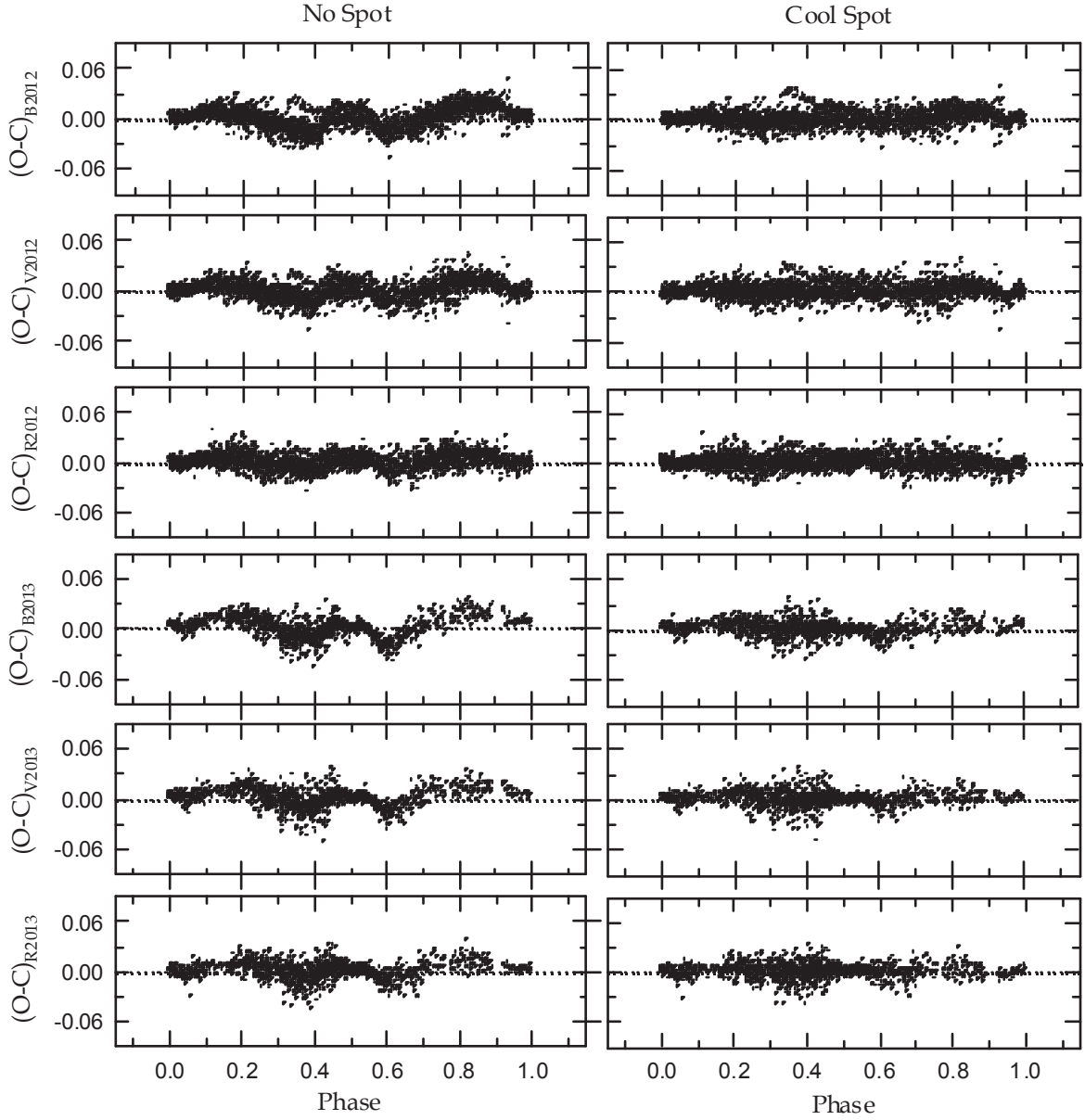}
\caption{Light residuals corresponding to two binary models in Table 4: without (left panels) and with (right panels) 
the inclusion of a cool spot on the primary component. }
\label{Fig4}
\end{figure}

\begin{figure}
\includegraphics[]{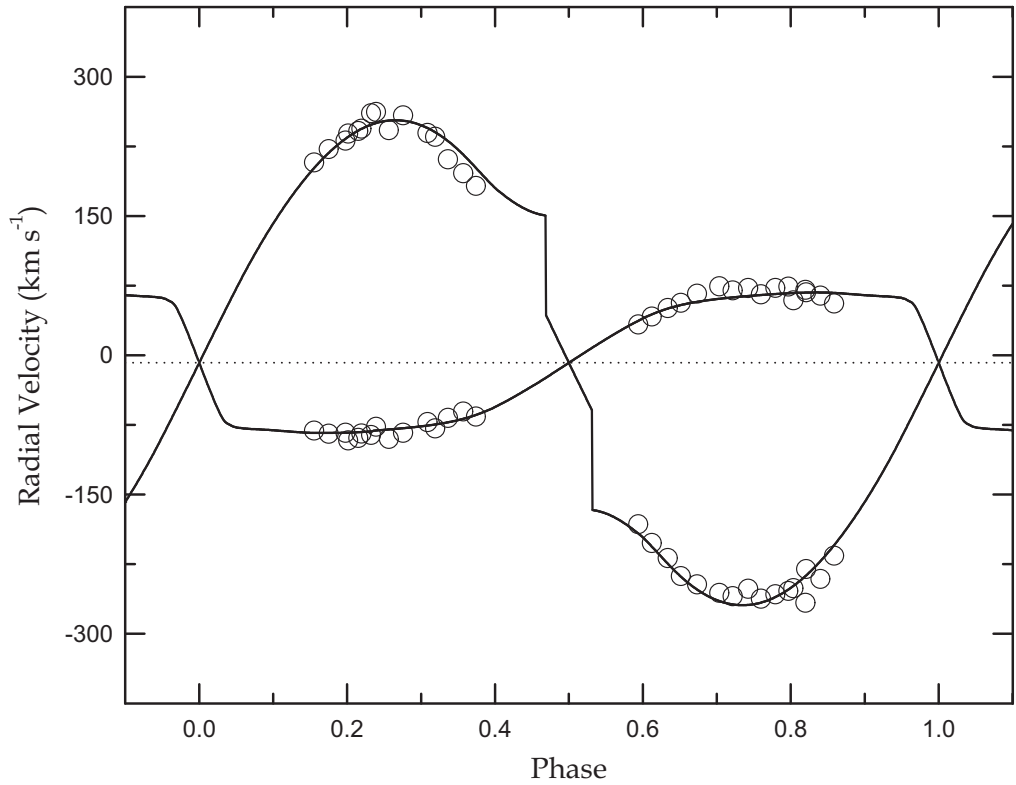}
\caption{Radial-velocity curves of UZ Leo. The open circles are the measures of Rucinski \& Lu (1999), while the solid curves 
denote the result from consistent light and velocity curve analysis including proximity effects. The dotted line refers to 
the systemic velocity of $-$8.0 km s$^{-1}$. }
\label{Fig5}
\end{figure}

\clearpage
\begin{deluxetable}{crcrcr}
\tabletypesize{\small}
\tablewidth{0pt} 
\tablecaption{CCD photometric data of UZ Leo observed in 2012 and 2013.}
\tablehead{
\colhead{HJD} & \colhead{$\Delta B$} & \colhead{HJD} & \colhead{$\Delta V$} & \colhead{HJD} & \colhead{$\Delta R$} 
}
\startdata
2,455,977.25010 & $-$2.3153  &  2,455,977.25039 & $-$2.2050  &  2,455,977.25066 & $-$2.1134   \\
2,455,977.25132 & $-$2.3214  &  2,455,977.25162 & $-$2.2067  &  2,455,977.25188 & $-$2.1101   \\
2,455,977.25229 & $-$2.3084  &  2,455,977.25259 & $-$2.1871  &  2,455,977.25287 & $-$2.1119   \\
2,455,977.25330 & $-$2.3155  &  2,455,977.25360 & $-$2.2008  &  2,455,977.25388 & $-$2.1073   \\
2,455,977.25431 & $-$2.3131  &  2,455,977.25462 & $-$2.1960  &  2,455,977.25490 & $-$2.1127   \\
2,455,977.25532 & $-$2.3193  &  2,455,977.25563 & $-$2.1875  &  2,455,977.25591 & $-$2.1173   \\
2,455,977.25634 & $-$2.3032  &  2,455,977.25664 & $-$2.1846  &  2,455,977.25693 & $-$2.1146   \\
2,455,977.25735 & $-$2.3085  &  2,455,977.25766 & $-$2.2011  &  2,455,977.25794 & $-$2.1144   \\
2,455,977.25837 & $-$2.3100  &  2,455,977.25969 & $-$2.2005  &  2,455,977.25896 & $-$2.1151   \\
2,455,977.25939 & $-$2.3205  &  2,455,977.26071 & $-$2.1885  &  2,455,977.25998 & $-$2.1173   \\
\enddata
\tablecomments{This table is available in its entirety in machine-readable and Virtual Observatory (VO) forms.}
\end{deluxetable}

\begin{deluxetable}{llrccl}
\tabletypesize{\small}
\tablewidth{0pt} 
\tablecaption{Observed photoelectric and CCD times of minimum light for UZ Leo.}
\tablehead{
\colhead{HJD} & Error & \colhead{Epoch} & \colhead{$O$--$C_{\rm full}$} & \colhead{Min} & \colhead{References}  \\
\colhead{(2,400,000+)} & & & & &  }
\startdata
34,041.4777   &               &  $-$9,318.0  &  $-$0.00092  &  I   &  Smith (1959)                          \\
34,478.4295   &               &  $-$8,611.0  &  $-$0.00159  &  I   &  Hinderer (1960)                       \\
34,486.4650   &               &  $-$8,598.0  &  $-$0.00057  &  I   &  van Houten (1956)                     \\
34,864.3967   &               &  $-$7,986.5  &  $+$0.00102  &  II  &  Hinderer (1960)                       \\
35,873.655    &               &  $-$6,353.5  &  $+$0.00274  &  II  &  Broglia \& Lenouvel (1960)            \\
38,855.6932   &               &  $-$1,528.5  &  $-$0.00258  &  II  &  Binnendijk (1972)                     \\
40,673.6666   &               &     1,413.0  &  $-$0.00114  &  I   &  Binnendijk (1972)                     \\
41,767.606    &               &     3,183.0  &  $+$0.00031  &  I   &  Strauss (1976)                        \\
44,638.4321   & $\pm$0.0002   &     7,828.0  &  $+$0.00165  &  I   &  Yang \& Liu (1982)                    \\
45,079.402    &               &     8,541.5  &  $-$0.00576  &  II  &  Diethelm (1982)                       \\
\enddata
\tablecomments{This table is available in its entirety in machine-readable and Virtual Observatory (VO) forms.}
\end{deluxetable}

\begin{deluxetable}{lcc}
\tablewidth{0pt}
\tablecaption{Parameters for the quadratic {\it plus} LTT ephemeris of UZ Leo. }
\tablehead{
\colhead{Parameter}     & \colhead{Value}                          & \colhead{Unit}}
\startdata                                
$T_0$                   & 2,439,800.37693$\pm$0.00063              & HJD             \\
$P$                     & 0.618044426$\pm$0.000000027              & d               \\
$a_{\rm b}\sin i_{3}$   & 3.94$\pm$0.12                            & au              \\
$e_{\rm b}$             & 0.19$\pm$0.14                            &                 \\
$\omega_{\rm b}$        & 137.0$\pm$4.6                            & deg             \\
$n_{\rm b}$             & 0.00710$\pm$0.00014                      & deg d$^{-1}$    \\
$T_{\rm b}$             & 2,484,158$\pm$764                        & HJD             \\
$P_{\rm b}$             & 138.8$\pm$2.8                            & yr              \\
$K_{\rm b}$             & 0.02252$\pm$0.00069                      & d               \\
$f(M_{3})$              & 0.00317$\pm$0.00012                      & M$_\odot$       \\ [1.0mm]
$M_{3} \sin i_{3}$      & 0.3007$\pm$0.0056                        & M$_\odot$       \\
$a_{3} \sin i_{3}$      & 34.42$\pm$0.32                           & au              \\
$e_3$                   & 0.19$\pm$0.14                            &                 \\
$\omega_{3}$            & 317.0$\pm$4.6                            & deg             \\
$P_3$                   & 138.8$\pm$2.8                            & yr              \\ [1.0mm]
$A$                     & +2.950($\pm$0.011)$\times 10^{-10}$      & d               \\
$dP$/$dt$               & +3.487($\pm$0.013)$\times 10^{-7}$       & d yr$^{-1}$     \\ [1.0mm]
$\sigma _{\rm all} ^a$  &  0.0128                                  & d               \\
$\sigma _{\rm pc} ^b$   &  0.0018                                  & d               \\
$\chi^2 _{\rm red}$     &  1.075                                   &                 \\
\enddata                                                     
\tablenotetext{a}{rms scatter of all residuals.}
\tablenotetext{b}{rms scatter of the PE and CCD residuals.}
\end{deluxetable}

\begin{deluxetable}{lccccc}
\tabletypesize{\scriptsize}  
\tablewidth{0pt}
\tablecaption{Light and RV parameters of UZ Leo.}
\tablehead{
\colhead{Parameter}                         & \multicolumn{2}{c}{Without spot}          && \multicolumn{2}{c}{With spot}             \\ [1.0mm] \cline{2-3} \cline{5-6} \\[-2.0ex]
                                            & \colhead{Primary} & \colhead{Secondary}   && \colhead{Primary} & \colhead{Secondary}
}
\startdata                                                                         
$T_0$ (HJD)                                 & \multicolumn{2}{c}{2,456,011.865669(42)}  && \multicolumn{2}{c}{2,456,011.866118(36)}  \\
$P$ (day)                                   & \multicolumn{2}{c}{0.61805954(11)}        && \multicolumn{2}{c}{0.61806015(9)}         \\
$i$ (deg)                                   & \multicolumn{2}{c}{89.56(32)}             && \multicolumn{2}{c}{87.35(18)}             \\
$T$ (K)                                     & 6980(250)         & 7094(250)             && 6980(250)         & 6772(250)             \\
$\Omega$                                    & 2.3375(17)        & 2.3375                && 2.3354(15)        & 2.3354                \\
$\Omega_{\rm in}$                           & \multicolumn{2}{c}{2.4823}                && \multicolumn{2}{c}{2.4802}                \\
$f$ (\%)                                    & \multicolumn{2}{c}{75.7}                  && \multicolumn{2}{c}{75.9}                  \\
$X$, $Y$                                    & 0.639, 0.254      & 0.639, 0.257          && 0.639, 0.254      & 0.638, 0.249          \\
$x_{B}$, $y_{B}$                            & 0.790, 0.274      & 0.788, 0.283          && 0.790, 0.274      & 0.795, 0.256          \\
$x_{V}$, $y_{V}$                            & 0.692, 0.289      & 0.688, 0.293          && 0.692, 0.289      & 0.698, 0.283          \\
$x_{R}$, $y_{R}$                            & 0.594, 0.294      & 0.589, 0.296          && 0.594, 0.294      & 0.603, 0.289          \\
$L_1$/($L_{1}$+$L_{2}$+$L_{3}$){$_{B2012}$} & 0.6272(26)        & 0.2616                && 0.6650(22)        & 0.2140                \\
$L_1$/($L_{1}$+$L_{2}$+$L_{3}$){$_{V2012}$} & 0.6483(25)        & 0.2642                && 0.6784(22)        & 0.2263                \\
$L_1$/($L_{1}$+$L_{2}$+$L_{3}$){$_{R2012}$} & 0.6662(23)        & 0.2666                && 0.6897(21)        & 0.2359                \\
$L_1$/($L_{1}$+$L_{2}$+$L_{3}$){$_{B2013}$} & 0.6264(27)        & 0.2612                && 0.6663(24)        & 0.2144                \\
$L_1$/($L_{1}$+$L_{2}$+$L_{3}$){$_{V2013}$} & 0.6461(26)        & 0.2632                && 0.6784(23)        & 0.2263                \\
$L_1$/($L_{1}$+$L_{2}$+$L_{3}$){$_{R2013}$} & 0.6637(25)        & 0.2657                && 0.6894(23)        & 0.2358                \\
{\it $l_{3B2012}$$\rm ^a$}                  & \multicolumn{2}{c}{0.1112(23)}            && \multicolumn{2}{c}{0.1210(18)}            \\
{\it $l_{3V2012}$$\rm ^a$}                  & \multicolumn{2}{c}{0.0875(22)}            && \multicolumn{2}{c}{0.0953(18)}            \\
{\it $l_{3R2012}$$\rm ^a$}                  & \multicolumn{2}{c}{0.0672(22)}            && \multicolumn{2}{c}{0.0744(18)}            \\
{\it $l_{3B2013}$$\rm ^a$}                  & \multicolumn{2}{c}{0.1124(25)}            && \multicolumn{2}{c}{0.1193(20)}            \\
{\it $l_{3V2013}$$\rm ^a$}                  & \multicolumn{2}{c}{0.0907(24)}            && \multicolumn{2}{c}{0.0953(20)}            \\
{\it $l_{3R2013}$$\rm ^a$}                  & \multicolumn{2}{c}{0.0706(24)}            && \multicolumn{2}{c}{0.0748(20)}            \\
$r$ (pole)                                  & 0.4852(4)         & 0.2976(7)             && 0.4855(4)         & 0.2974(7)             \\
$r$ (side)                                  & 0.5315(6)         & 0.3160(10)            && 0.5319(5)         & 0.3159(9)             \\
$r$ (back)                                  & 0.5701(9)         & 0.3957(34)            && 0.5704(8)         & 0.3957(32)            \\
$r$ (volume)$\rm ^b$                        & 0.5304(6)         & 0.3322(13)            && 0.5307(5)         & 0.3320(13)            \\ 
Colatitude (deg)                            & \dots             & \dots                 && 50.07(8)          & \dots                 \\
Longitude (deg)                             & \dots             & \dots                 && 184.4(2)          & \dots                 \\
Radius (deg)                                & \dots             & \dots                 && 31.17(6)          & \dots                 \\
$T$$\rm _{spot}$/$T$$\rm _{local}$          & \dots             & \dots                 && 0.927(1)          & \dots                 \\
$\Sigma W(O-C)^2$                           & \multicolumn{2}{c}{0.0136}                && \multicolumn{2}{c}{0.0110}                \\[1.0mm]
\multicolumn{6}{l}{Spectroscopic orbits:}                                                                                            \\ 
$T_0$ (HJD)                                 & \multicolumn{2}{c}{2,450,595.8339(41)}    && \multicolumn{2}{c}{2,450,595.8339(42)}    \\
$P$ (day)                                   & \multicolumn{2}{c}{0.618089(34)}          && \multicolumn{2}{c}{0.618089(34)}          \\
$\gamma$ (km s$^{-1}$)                      & \multicolumn{2}{c}{$-$7.9(1.0)}           && \multicolumn{2}{c}{$-$8.0(1.0)}           \\
$a$ (R$_\odot$)                             & \multicolumn{2}{c}{4.197(33)}             && \multicolumn{2}{c}{4.213(33)}             \\
$q$ $(=M_2/M_1)$                            & \multicolumn{2}{c}{0.3073(47)}            && \multicolumn{2}{c}{0.3063(49)}            \\
$\Sigma W(O-C)^2$                           & \multicolumn{2}{c}{0.0098}                && \multicolumn{2}{c}{0.0098}                  
\enddata
\tablenotetext{a}{Value at 0.75 phase. }
\tablenotetext{b}{Mean volume radius. }
\end{deluxetable}

\begin{deluxetable}{lcc}
\tablewidth{0pt}
\tablecaption{Absolute parameters for UZ Leo.}
\tablehead{
\colhead{Parameter}    & \colhead{Primary}   & \colhead{Secondary}}
\startdata                                    
$M$ (M$_\odot$)        & 2.01$\pm$0.03       & 0.62$\pm$0.01          \\
$R$ (R$_\odot$)        & 2.23$\pm$0.01       & 1.40$\pm$0.01          \\
$\log$ $g$ (cgs)       & 4.04$\pm$0.01       & 3.94$\pm$0.01          \\
$\rho$ (g cm$^3)$      & 0.255$\pm$0.006     & 0.319$\pm$0.010        \\
$L$ (L$_\odot$)        & 10.6$\pm$1.5        & 3.68$\pm$0.55          \\
$M_{\rm bol}$ (mag)    & $+$2.16$\pm$0.16    & $+$3.31$\pm$0.16       \\
BC (mag)               & $+$0.03$\pm$0.01    & $+$0.02$\pm$0.01       \\
$M_{\rm V}$ (mag)      & $+$2.13$\pm$0.16    & $+$3.29$\pm$0.16       \\
Distance (pc)          & \multicolumn{2}{c}{388$\pm$29}                
\enddata                                                     
\end{deluxetable}

\end{document}